# Analysis and Mitigation of Cascading Failures Using a Stochastic Interaction Graph with Eigen-analysis

Zhenping Guo, *Student Member, IEEE*, Xiaowen Su, Kai Sun, *Fellow, IEEE*,
Byungkwon Park, *Member, IEEE,* Srdjan Simunovic

*Abstract* – In studies on complex network systems using graph theory, eigen-analysis is typically performed on an undirected graph model of the network. However, when analyzing cascading failures in a power system, the interactions among failures suggest the need for a directed graph beyond the topology of the power system to model directions of failure propagation. To accurately quantify failure interactions for effective mitigation strategies, this paper proposes a stochastic interaction graph model and associated eigen-analysis. Different types of modes on failure propagations are defined and characterized by the eigenvalues of a stochastic interaction matrix, whose absolute values are unity, zero, or in between. Finding and interpreting these modes helps identify the probable patterns of failure propagation, either local or widespread, and the participating components based on eigenvectors. Then, by lowering the failure probabilities of critical components highly participating in a mode of widespread failures, cascading can be mitigated. The validity of the proposed stochastic interaction graph model, eigen-analysis and the resulting mitigation strategies is demonstrated using simulated cascading failure data on an NPCC 140-bus system.

## I. Introduction

CASCADING failures often occur in complex systems, such as power systems, Internet, and interdependent infrastructure systems. In power systems, cascading failures are complicated sequences of interdependent outages which could result in detrimental economic and social effects [1], [2]. For example, the 2003 Northeast blackout [3], the 2011 Arizona-Southern California blackout [4], the 2012 Indian blackout [5], the 2019 Argentina, Paraguay, and Uruguay blackout [6]. To effectively prevent or mitigate cascading failures, it is essential to accurately quantify the interactions among interdependent failures and study the propagation mechanisms involved.

Cascading failure analyses and studies are usually conducted on two types of models: physical models [7]-[13] and probabilistic models. Physical models are constructed based on physical laws of power systems to simulate cascading failure sequences. These models can capture detailed power system dynamics of cascading failures, but they are mainly used for offline studies due to the heavy computational burdens and the limited time available in the real-time operating environment for operators' decisions. In contrast, probabilistic models ignore the power system dynamics, and focus on finding the characteristics and patterns of failure propagation. Typical models are such as the interaction graph models [14]-[21] and influence graph models [22]-[24]. Such models can be constructed offline by data-driven approaches using historical or simulated cascade event data, quantifying stochastic interactions between interdependent failures in power systems. The key components of cascading failures can be efficiently derived from these models, providing valuable insights for real-time control to prevent or mitigate cascading failures.

In a sequence of cascading failures, multiple component failures that occur closely in time are often grouped into the same generation for easier handling. Thus, all failures in the sequence are partitioned into a number of generations ordered in time. In most interaction graph models, each node of the graph represents a single component of the power system. Thus, to construct the graph with edges connecting individual nodes, one-to-one causal relationships often need to be inferred or assumed between multiple component failures respectively from two successive generations. For example, [14] assumes that a component failure is caused by the most frequently failed component in the previous generation in the entire failure dataset. For a more efficient estimation of the interactions among multiple component failures, [16] estimates one-to-one causal relationships between successive generations using an expectation maximization algorithm. To identify critical components for upgrades and mitigate the risk of cascading failures, [22] evenly distributes the impact among multiple failed lines of the same generation to construct an influence graph on failure propagations. Using real outage data, [17] estimates the causal relationships considering more than two successive generations. Ref. [23] treats nearly simultaneous component failures as one single state and adds an absorbing state ending the propagation, resulting in a well-defined Markovian influence graph. The estimation of transition probabilities distinguishes the first transition from subsequent ones for a better match with the statistics on failure propagations in real data. Furthermore, it applies eigen-analysis to identify the transmission lines that highly participate in more severe cascading scenarios, leveraging the asymptotic properties of the Markov chain. There are some other studies using Markov chains to model cascading failures [25]-[26]. A Markov-transition model based on power flow redistribution is proposed in [25] to capture the evolution of cascading failures and identify the most probable propagation path. A continuous-time Markov chain is utilized in [26] to model the dynamics

This work was supported by UT-Battelle, LLC, under Contract No. DE-AC0500OR22725 with the U.S. Department of Energy.

Z. Guo, X. Su and K. Sun are with the Department of EECS, University of Tennessee, Knoxville, TN, USA (e-mail: zguo19@vols.utk.edu, suxiaowen0508@gmail.com, kaisun@utk.edu).

B. Park is with the Department of Electrical Engineering, Soongsil University, Seoul 06978, South Korea (e-mail: bkpark@ssu.ac.kr).

S. Simunovic is with the Computational Sciences and Engineering Division, Oak Ridge National Laboratory, Oak Ridge, TN, 37830, USA (e-mail: simunovics@ornl.gov).

with cascading failures, enabling discovering the key operating characteristics that may cause severe cascading failures.

This paper proposes a new, state-based stochastic interaction graph model also based on assumptions on Markov chains. The model estimates failure interactions based on empirical conditional probabilities over a historical or simulated cascading failure dataset. Unlike the estimation methods in [17] and [23], in which the probabilities of both first and subsequent generations are considered but handled differently, this paper ignores scenarios having only one generation of failures and focuses on those causing subsequent failures. Thus, the resulting stochastic interaction graph will not focus on early transitions of failures. The model is a directed graph with weighted edges and self-loops, where component failures occurring close in time are considered as one state of failure propagation and is defined as a vertex in the graph. Compared with a component-based interaction graph in, e.g., [14], this model avoids misinterpretation of interactions between component failures when high resolution data are unavailable. Additionally, to distinguish failure propagations that end at different states, self-loops are defined to represent ending failures in the proposed graph without introducing an extra vertex [23]. Furthermore, the paper identifies the characteristics of failure propagation beyond the physical topology of the power system by using eigen-analysis on the stochastic interaction model. The eigen-analysis reveals different types of modes of failure propagation, which are persistent modes, trivial modes, and transient modes characterized by eigenvalues, whose absolute values are respectively unity, zero, and in between. These modes provide insights into different characteristics of interest and can guide the development of effective mitigation strategies against cascading failures.

In studies on power systems and other complex network systems using graph theory [27], eigen-analysis is typically performed on an undirected graph model that represents the network's topology [28]. Some theories on spectrum analysis of graphs are available to understand the characteristics of the physical topology of a power system as an undirected graph [29]. However, the sequences, interactions, and causalities of failures in a power system are usually directed and hence are better modeled by directed graphs, which are not its physical topology [30], [31]. Some studies on eigen-analysis of directed graphs discovered new information unavailable from eigen-analysis of undirected graphs. For example, an undirected graph modeled by a real symmetric matrix has all real eigenvalues, which is not necessarily true for a directed graph. Eigen-analysis utilized to understand modal properties of a square matrix defining a transformation [32], can be borrowed to study the modal properties of a directed graph modeling failure interactions in a power system. In this paper, the directed interaction graph is modeled based on assumptions of Markov chains. There are multiple studies on the eigen-analysis of stochastic matrices of Markov chains representing directed graphs [33]-[35]. The stochastic matrix of an irreducible Markov chain has a unity eigenvalue. For a decomposable stochastic matrix, which is reducible, its eigenvectors corresponding to unity eigenvalues can be utilized to distinguish essential and transient states [33]. The eigenvector corresponding to the subdominant eigenvalue of an irreducible stochastic matrix can indicate state clustering. Several methods have been developed for state clustering of Markov chains using one or more eigenvalues [34]-[40].

In this paper, we apply eigen-analysis to the stochastic matrix of the proposed directed interaction graph and interpret each mode consisting of its eigenvalue and eigenvector to reveal different patterns of failure propagation. This allows the development of effective mitigation strategies against cascading failures based on the modal information. The computational complexity of the proposed approach depends on the selected subsets of failure components in the data used to construct the model, which makes the dimensions of the stochastic matrix adjustable and hence enables the scalability of this approach for analyzing and mitigating cascading failures in large-scale power systems.

In the rest of the paper, Section II discusses how the interactions between interdependent failures are quantified using a state-based stochastic interaction graph. Section III conducts the eigen-analysis on this graph and classifies its modes. In Section IV, the proposed stochastic interaction graph model and eigen-analysis are validated using cascading failure data simulated from an NPCC (Northeast Power Coordinating Council) 140-bus system. Different mitigation strategies are applied and compared, showing the importance of modal information in analysis and mitigation of cascading failures. Section V draws conclusions of the paper.

## II. PROPOSED STOCHASTIC INTERACTION GRAPH

This section introduces a novel state-based stochastic interaction graph model that has been quantified using cascading failure data. By comparing the proposed state-based stochastic interaction model with an existing component-based interaction model, the advantages of the former are highlighted.

### A. Cascading Failure Data

Cascading failure data can be obtained from historical events or cascading failure simulations. When using historical events, the component failures that happen closely in time, e.g. within one hour [41], are often grouped into a cascading failure scenario called a "cascade", which describes a sequence of failures starting from an initial failure. Further in each cascade, the failures that happened very closely in time, e.g. within one minute [41], can be considered concurrent and grouped into one generation with interdependency ignored. Thus, each cascade can have a number of generations until the end of cascading failures. When real data of cascading failures are unavailable or insufficient, cascades can also be generated using a cascading failure simulator on a power system model starting from assumed initial failures. The simulated failures in each cascade can be organized similarly in generations.

To analyze the interactions between interdependent failures, a large number $M$ of cascades from the data are arranged in generations 0, 1, 2, …, as shown in Fig. 1 [14], where $F_g^{(m)}$ represents the set of failed components in the $g$-th generation of cascade $m$. The $0^{th}$ generation consists of initial failures. For instance, if $F_2^{(3)}=\{c_1, c_3, c_7\}$, the failures of the $1^{st}$, $3^{rd}$, and $7^{th}$ components are recorded in the 2nd generation of cascade 3. Each cascade ends after a finite number of generations. Note that this paper excludes the cascades stopped at generation 0 and focuses on more severe cascades whose initial failures can cause subsequent failures, ensuring that each cascade concludes after at least two generations.

| | generation 0 | generation 1 | generation 2 | ... |
|---|---|---|---|---|
| cascade 1 | $F_0^{(1)}$ | $F_1^{(1)}$ | $F_2^{(1)}$ | ... |
| cascade 2 | $F_0^{(2)}$ | $F_1^{(2)}$ | $F_2^{(2)}$ | ... |
| ⋮ | ⋮ | ⋮ | ⋮ | ⋮ |
| cascade $M$ | $F_0^{(M)}$ | $F_1^{(M)}$ | $F_2^{(M)}$ | ... |

Fig. 1. Cascading failure data organized in cascades and generations

### B. State-based Stochastic Interaction Graph Model

The proposed graph model is a directed graph with weighted edges $G(V, E, W)$ consisting of a vertex set $V=\{s_1, s_2, ..., s_N\}$ whose cardinality is $N$ (i.e., $|V|=N$), an edge set $E \subset V \times V$ where an edge connects an ordered pair of vertices of $V$, and a weight set $W$. Each vertex $s \in V$ representing one state of failure propagation is defined as the entire set of failed components recorded by any one generation of a cascade. A directed edge $e_{ij}=(s_i, s_j) \in E$ from source vertex $s_i$ to sink vertex $s_j$ represents a probable failure propagation by one generation from the failure of vertex $s_i$ to the failure of vertex $s_j$. Each edge $e_{ij}$ is assigned a weight $w_{ji} \in W$ defined and estimated by (1) to model the conditional probability of vertex $s_j$ failing immediately after vertex $s_i$ fails.

$$w_{ji} = \Pr\{X_{g+1}=s_j \mid X_g=s_i\} \approx n_{ji}/\sum_j n_{ji} \quad (1)$$

where $n_{ji}$ counts how many times vertex $s_j$ fails immediately after vertex $s_i$ across the entire dataset. The estimation of $w_{ji}$ approaches the actual probability with enough data. There should be $\Sigma_j w_{ji}=1$, $s_i \in V$; i.e., the weight sum for all edges leaving from a vertex equals one.

If vertex $s_i$ failing does not cause any other to fail over the cascade dataset, it is considered an absorbing state [33] with $n_{ji}=0$ for all $j$. To avoid a division-by-zero error in (1), add a self-loop $e_{ii}=(s_i, s_i) \in E$ to each absorbing state $s_i$ of weight one:

$$w_{ii} = \Pr\{X_{g+1}=s_i \mid X_g=s_i\} = 1 \quad (2)$$

In [23], a common vertex is introduced in the Markov influence graph that represents all ending failures. In contrast, the introduction of self-loops to absorbing states allows for the distinction of failure propagations that end at different states.

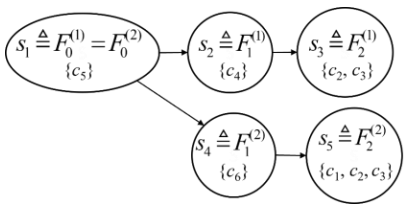

Fig. 2. Illustration on distinct vertices and directed edges.

As illustrated by Fig. 2, if first two cascades involve failures of components $c_1$ to $c_6$ in two generations: $F_0^{(1)}=\{c_5\}$, $F_1^{(1)}=\{c_4\}$, $F_2^{(1)}=\{c_2, c_3\}$, $F_0^{(2)}=\{c_5\}$, $F_1^{(2)}=\{c_6\}$, and $F_2^{(2)}=\{c_1, c_2, c_3\}$. The interaction graph may have five distinct vertices, i.e., $s_1$, $s_2$, $s_3$, $s_4$, and $s_5$, and four directed edges, i.e., $e_{12}$, $e_{23}$, $e_{14}$, and $e_{45}$. Note that vertex $s_1$ corresponds to $F_0^{(1)}=F_0^{(2)}$ while the other four states are defined by four different sets, respectively.

### C. Component-based and State-based Interaction Graphs

To demonstrate the superiority of the proposed state-based stochastic interaction graph, it is compared with a component-based interaction graph for a system having failures of three lines as shown in Fig. 3:

| | generation 0 $X_0$ | generation 1 $X_1$ | generation 2 $X_2$ | ... |
|---|---|---|---|---|
| cascade 1 | $s_1$ = {line 1} | $s_3$ = {line 3} | $s_2$ = {line 2} | {} |
| cascade 2 | $s_2$ = {line 2} | $s_4$ = {line 1, line 3} | {} | ... |
| cascade 3 | $s_2$ = {line 2} | $s_1$ = {line 1} | {} | ... |
| cascade 4 | $s_3$ = {line 3} | $s_2$ = {line 2} | {} | ... |

(a)

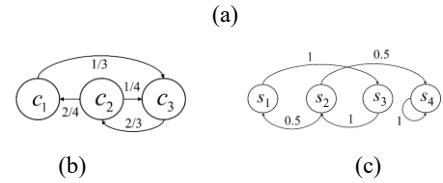

(b)                         (c)

Fig. 3. Comparing component-based and state-based interaction graphs.

In a component-based interaction graph, a vertex typically represents the failure of a single component, such as a transformer or line, and interactions are assumed between any two components from two consecutive generations of a cascade. The edges and their weights in the graph can quantify empirical probabilities of failure interactions, but this requires information on interactions between individual components. However, such detail is usually unavailable from raw data due to insufficient time resolution. In reality, component failures that occur closely in time are often grouped into the same generation or stage and then treated as concurrent failures tagged with the same time. As a result, this coarse handling may cause misinterpreting interactions between failures. The quantification of a component-based interaction graph is illustrated in Fig. 3 (b), which models failures of three components $c_1$, $c_2$, $c_3$ linked by four directed edges $e_{13}=(c_1, c_3)$, $e_{21}=(c_2, c_1)$, $e_{23}=(c_2, c_3)$, $e_{32}=(c_3, c_2)$. The weight of each edge $w_{ji}$ is estimated by $b_{ij}$ in [14] as $w_{31}=1/3$, $w_{12}=2/4$, $w_{32}=1/4$, $w_{23}=2/3$. Notably, the existence of edge $e_{23}$ indicates that the 3rd line may fail right after the 2nd line fails, which, however, cannot be observed directly from the cascades in Fig. 3 (a). Rather, either the 1st line fails or the 1st and 3rd lines both fail right after the 2nd line fails in the data. From this case, a component-based interaction graph attempts to quantify failure interactions down to the component level, which may not be evidenced from the data only verified at the generation level.

To address this issue, the proposed state-based stochastic interaction graph aims to more reliably quantify failures interactions at the generation level to avoid misinterpretation from the graph. Here, a vertex represents the entire set of failure components recorded in one generation of a cascade and an interaction is assumed between any two consecutive generations (i.e., states) of a cascade. Since the vertex can represent different states of failure propagation, such as the failure of one component or the concurrent failures of multiple components from the existing data, the proposed graph ensures the correct identification of edges between any two states, thereby ensuring accurate estimation of conditional probabilities of failure interactions from the available data.

As illustrated in Fig. 3 (c), consider four vertices representing four failure propagation states: $s_1$, $s_2$, and $s_3$ represent respectively the failures of lines 1, 2, and 3, and $s_4$ represents the concurrent failures of lines 1 and 3. The graph has five directed edges: $e_{13}=(s_1, s_3)$, $e_{32}=(s_3, s_2)$, $e_{24}=(s_2, s_4)$, $e_{21}=(s_2, s_1)$, $e_{44}=(s_4, s_4)$ based on the data in Fig. 3 (a). Edge weights are calculated by (1) and (2) as $w_{31}=1$, $w_{23}=1$, $w_{42}=0.5$, $w_{12}=0.5$, and $w_{44}=1$. Self-loops are only assigned to absorbing states that do not cause any subsequent failures over the entire dataset. Here only $s_4$ is an absorbing state with a self-loop because it always concludes a cascade if it fails according to the entire dataset. All edges are inferred from direct observations of the data by checking possible interactions between two consecutive generations. For example, edge $e_{24}$ with weight $w_{42}$ shown in the state-based interaction graph represents one probable step of failure propagation from line 2 to the concurrent failures of lines 1 and 3. In contrast, the component-based interaction graph in Fig. 3 (b) does not depict the concurrent failures of $s_4$, and hence may not accurately estimate its conditional probability when such concurrent failures need to be analyzed or mitigated.

### D. State-based Stochastic Interaction Matrix

An $N$-vertex stochastic interaction graph $G$ ($V$, $E$, $W$) can be characterized by an $N \times N$ stochastic interaction matrix $\mathbf{W}$ whose $j$-th row $i$-th column entry is either zero or equal to weight $w_{ji} \in W$ of directed edge $e_{ij}$ from vertex $s_i$ to vertex $s_j$. The stochastic matrix of the interaction graph describes the transitions of a Markov chain, assuming that the failures in the next generation rely solely on the failures in the current generation. This assumption enables the analysis of the propagation of cascading failures in an analytically tractable manner. Without loss of generality, the stochastic matrix in this paper is defined as a left stochastic matrix for the Markov chain, in which each column sums to one [42]. (Equivalently, the stochastic matrix can also be defined as a right stochastic matrix in which each row sums to one [33].) $\mathbf{W}$ is a non-negative real square matrix and can be transformed into its Jordan Canonical form [32], which is an eigen-decomposed block diagonal matrix with $P$ Jordan blocks.

$$\mathbf{J} = \mathbf{V}^{-1}\mathbf{W}\mathbf{V} = diag(\mathbf{J}_1, \ldots, \mathbf{J}_P) \in \mathbf{C}^{N \times N} \quad (3)$$

Each block $\mathbf{J}_i$ has a dimension $N_i \geq 1$, and there is $N=N_1+N_2+\ldots+N_P$. Arrange $\mathbf{J}$ to have first $q$ Jordan blocks to be 1-dimensional, i.e. $\mathbf{J}_i = \lambda_i$ for $i=1, \ldots, q$, and the remaining ($P-q$) Jordan blocks to be multi-dimensional for $i=q+1, \ldots, P$. $\mathbf{V}=\{\mathbf{v}_1, \ldots, \mathbf{v}_N\} \in \mathbf{C}^{N \times N}$ is made of normalized, generalized right eigenvectors satisfying $\|\mathbf{v}_i\|_\infty=1$ ($i=1, 2, \ldots, N$). $v_{ji}$ as the $j$-th entry of $\mathbf{v}_i$ reveals vertex $s_j$'s participation in the $i$-th mode as shown later.

The distribution of failure probabilities on $N$ states (i.e. vertices) in generation $k$ is represented by an $N$-dimensional non-negative real vector $\mathbf{p}^{(k)}=[p_1^{(k)}, p_1^{(k)}, \ldots, p_N^{(k)}]^T$, ($k \in \mathbf{Z}^+$) whose sum equals 1. Particularly, $p_i^{(k)}$ is the failure probability of $s_i$ in generation $k$. Use a complex vector $\mathbf{c}^{(k)}=[c_1^{(k)}, c_2^{(k)}, \ldots, c_N^{(k)}]^T=\mathbf{V}^{-1}\mathbf{p}^{(k)}$ to denote $\mathbf{p}^{(k)}$'s coordinates in the eigenspace.

In the rest of the paper, eigen-analysis is performed on the stochastic matrix $\mathbf{W}$, and its modes including eigenvalues and eigenvectors are identified and interpreted to reveal characteristics and patterns of the stochastic process associated with cascading failures.

## III. EIGEN-ANALYSIS OF STOCHASTIC INTERACTION GRAPH

### A. Definition of a Mode

At the last generation $K$,

$$\mathbf{p}^{(K)} = \mathbf{W}^K \mathbf{p}^{(0)} = \mathbf{V}\mathbf{J}^K \mathbf{c}^{(0)} = \sum_{i=1}^{q} c_i^{(0)} \lambda_i^K \mathbf{v}_i + \sum_{i=q+1}^{P} \left[ \mathbf{v}_i, \mathbf{v}_{i+1}, \cdots \mathbf{v}_{i+N_i-1} \right] \mathbf{J}_i^K \begin{bmatrix} c_i^{(0)} \\ c_{i+1}^{(0)} \\ \vdots \\ c_{i+N_i-1}^{(0)} \end{bmatrix} \quad (4a)$$

$$\mathbf{J}_i^K = \begin{bmatrix} \lambda_i^K & \binom{K}{1}\lambda_i^{K-1} & \cdots & \binom{K}{N_i-1}\lambda_i^{K-N_i+1} \\ & \lambda_i^K & \ddots & \vdots \\ & & \ddots & \binom{K}{1}\lambda_i^{K-1} \\ & & & \lambda_i^K \end{bmatrix} \quad (4b)$$

where $\mathbf{J}_i^K$ is the $K$-th power of the $i$-th multi-dimensional Jordan block with a dimension of $N_i$ ($i \in [q+1, \ldots, P]$) as shown in (4b). In this study, the Jordan block with a zero eigenvalue is observed to be the only multi-dimensional Jordan block, whose generalized eigenvectors $\{\mathbf{v}_i, \mathbf{v}_{(i+1)}, \ldots, \mathbf{v}_{(i+N_i-1)}\}$ can be hard to obtain in the programming, while its effect is negligible after $K$ generations because $\mathbf{J}_i^K = \mathbf{0}$ when $K \geq N_i$ as indicated in (4b). Therefore, this multi-dimensional Jordan block can be approximated by multiple 1-dimensional Jordan blocks with a formula, i.e., $\|\mathbf{W}\mathbf{v}_j\text{-}\mathbf{0}\| \to 0$ ($j \in [i, \ldots, i+N_i-1]$), which are defined as trivial modes by **Definition 3** and discussed later.

Different types of failure propagation modes may exist. In the following, the eigenvalues and corresponding modes are categorized according to [39], and the influence of each mode is interpreted and discussed below.

**Definition 1 (mode)**: A *mode* of failure propagation is an ordered pair ($\lambda_i$, $\mathbf{v}_i$) satisfying $\mathbf{W}\mathbf{v}_i=\lambda_i\mathbf{v}_i$, $\mathbf{v}_i \neq \mathbf{0}$.

The $\mathbf{v}_i$ defines the mode shape of the mode. If $\lambda_i \in \mathbf{C}$, its conjugate ($\bar{\lambda}_i$, $\bar{\mathbf{v}}_i$) is also a mode with $\angle\bar{\lambda}_i = -\angle\lambda_i$. For any mode of $\mathbf{W}$ satisfying $\lambda_i \in \mathbf{R}$, there exists an eigenvector $\mathbf{v}_i \in \mathbf{R}^{N \times 1}$ [42].

### B. Persistent Mode ($\lambda_i=1$)

If $|\lambda_i|=1$, it indicates a persistent mode of failure propagation, whose failure probabilities never decrease. Considering two facts: first, a cascade propagating on the power grid must have a finite number of generations, and second, a failed component is rarely recovered on its own during the cascade, we can conclude that a persistent mode with $|\lambda_i|=1$ only exists at an absorbing state as defined below having $\lambda_i \equiv 1$.

**Definition 2 (persistent mode):** A *persistent mode* is a mode that is associated with an absorbing state $s_k$ having a self-loop and satisfies ($\lambda_i$, $\mathbf{v}_i$) ≡ (1, $\mathbf{e}_k$), where $\mathbf{e}_k$ is the $k$-th column of the identity matrix.

**Theorem 1:** For a persistent mode $(1, \mathbf{e}_k)$, if $\mathbf{p}^{(0)} = \mathbf{v}_i = \mathbf{e}_k$, then $\mathbf{p}^{(K)} = \mathbf{v}_i$ for any $K$.

*Proof:*

$$\begin{aligned}\mathbf{p}^{(K)} &= \mathbf{W}^K \mathbf{p}^{(0)} = \mathbf{W}^K \mathbf{v}_i = \mathbf{W}^{K-1} \mathbf{W} \mathbf{v}_i \\ &= \mathbf{W}^{K-1} \lambda_i \mathbf{v}_i = \lambda_i^K \mathbf{v}_i = \mathbf{v}_i\end{aligned} \quad (5)$$

∎

Under a persistent mode, the failure of an absorbing state $s_k$ remains and it will not propagate to others. Note that this paper excludes the cascades whose initial failure does not cause any subsequent failure, so the persistent mode cannot be applied to the initial generation since absorbing states are all sinks in the interaction graph proposed in this paper.

### C. Trivial Mode ($\lambda_i = 0$)

If $\lambda_i = 0$, it indicates a failure propagation pattern that theoretically only lasts for at most one generation, which is defined as a trivial mode:

**Definition 3 (trivial mode):** A *trivial mode* is a mode $(\lambda_i, \mathbf{v}_i)$ satisfying $\lambda_i = 0$ and $\mathbf{W}\mathbf{v}_i = 0$.

The influence of a trivial mode $(\lambda_i, \mathbf{v}_i)$ on failure propagation at generation $K$ is $c_i^{(0)} \lambda_i^K \mathbf{v}_i$, which equals zero for $K \geq 1$. This property shows that the mode shape of a trivial mode represents a probable step of failure propagation that is unlikely to continue after one generation, so trivial modes indicate non-propagation characteristics of the stochastic interaction graph.

### D. Transient Mode ($0 < |\lambda_i| < 1$)

If $0 < |\lambda_i| < 1$, it indicates a decaying failure propagation mode, under which failures continue with decreasing probabilities.

**Definition 4 (transient mode):** A *transient mode* is a mode $(\lambda_i, \mathbf{v}_i)$ satisfying $0 < |\lambda_i| < 1$.

In general, transient modes can be categorized as two sub-types: transient modes with positive less-than-one eigenvalues, and transient modes with complex eigenvalues having nonzero angles. Note that a special case of the latter is a transient mode with a negative eigenvalue when the angle is $\pi$. $|\lambda_i|$ measures the decaying magnitude. Similarly, $|\lambda_i|^k$ can measure the decaying magnitude after $k$ generations.

#### 1) Positive Less-than-one Eigenvalue

A mode $(\lambda_i, \mathbf{v}_i)$ with a positive less-than-one eigenvalue is a transient mode satisfying $0 < \lambda_i < 1$ and $\mathbf{v}_i \in \mathbf{R}^{N \times 1}$.

#### 2) Complex Nonzero-angle Eigenvalue

A mode $(\lambda_i, \mathbf{v}_i)$ with a complex eigenvalue with a nonzero angle is a transient mode satisfying either $\angle \lambda_i = \pi$, $-1 < \lambda_i < 0$ and $\mathbf{v}_i \in \mathbf{R}^{N \times 1}$ or $\angle \lambda_i \neq \pi$, $\lambda_i \notin \mathbf{R}$, and $\mathbf{v}_i \in \mathbf{C}^{N \times 1}$.

Each transient mode with $\angle \lambda_i = \pi$ involves two vertices forming a 2-cycle. Note that a 2-cycle does not indicate an oscillation between the failures of two states; instead, it only means that the failures of two correlated states can vary their order within the dataset of cascades.

Each transient mode $(\lambda_i, \mathbf{v}_i)$ with $\angle \lambda_i \neq \pi$, $\lambda_i \notin \mathbf{R}$, and $\mathbf{v}_i \in \mathbf{C}^{N \times 1}$, has a corresponding conjugate mode $(\bar{\lambda}_i, \bar{\mathbf{v}}_i)$, which is a transient mode carrying the same failure propagation information as $(\lambda_i, \mathbf{v}_i)$. $|v_{ji}|$ measures the participation of the vertex $s_j$ in the mode $i$. The angle difference $\angle v_{ui} - \angle v_{ji} \in (0, \pi)$ measures the alignment of the vertex $s_u$ and the vertex $s_j$ in this mode. Each mode involves multiple vertices to form a subgraph. $\angle \lambda_i$, which is neither 0 nor $\pi$, will add an angle to the propagation by one generation. The overall decaying magnitude after $K$ generations is $|\lambda_i|^K$.

Let $(\lambda_i, \mathbf{v}_i)$ and $(\lambda_{i+1}, \mathbf{v}_{i+1})$ be a conjugated pair of transient modes, and $c_i^{(0)}$ and $c_{i+1}^{(0)}$ be the coordinates of any $\mathbf{p}^{(0)}$ projected on $\mathbf{v}_i$ and $\mathbf{v}_{i+1}$ where $c_i^{(0)} = \overline{c_{i+1}^{(0)}}$. Their influence at generation $K$ is

$$c_i^{(0)} \lambda_i^K \mathbf{v}_i + c_{i+1}^{(0)} \lambda_{i+1}^K \mathbf{v}_{i+1} = c_i^{(0)} \lambda_i^K \mathbf{v}_i + \overline{c_i^{(0)}} \bar{\lambda}_i^K \bar{\mathbf{v}}_i$$

$$= \left| c_i^{(0)} \right| e^{j \angle c_i^{(0)}} \left| \lambda_i \right|^K e^{jK\angle \lambda_i} \begin{bmatrix} |v_{1i}| e^{j \angle v_{1i}} \\ \vdots \\ |v_{Ni}| e^{j \angle v_{Ni}} \end{bmatrix} + \left| c_i^{(0)} \right| e^{-j \angle c_i^{(0)}} \left| \lambda_i \right|^K e^{-jK\angle \lambda_i} \begin{bmatrix} |v_{1i}| e^{-j \angle v_{1i}} \\ \vdots \\ |v_{Ni}| e^{-j \angle v_{Ni}} \end{bmatrix}$$

$$= \left| c_i^{(0)} \right| \left| \lambda_i \right|^K \begin{bmatrix} |v_{1i}| \left( e^{j \theta_{1i}} + e^{-j \theta_{1i}} \right) \\ \vdots \\ |v_{Ni}| \left( e^{j \theta_{Ni}} + e^{-j \theta_{Ni}} \right) \end{bmatrix} = 2 \left| c_i^{(0)} \right| \left| \lambda_i \right|^K \begin{bmatrix} |v_{1i}| \cos \theta_{1i} \\ \vdots \\ |v_{Ni}| \cos \theta_{Ni} \end{bmatrix}$$

(6a)

$$\theta_{ni} = K \angle \lambda_i + \angle v_{ni} + \angle c_i^{(0)}, \quad n = 1, 2, \cdots N \quad (6b)$$

The product of $K$ and $\angle \lambda_i$ determines the total phasing change after a $K$-generation failure propagation. $2\pi / \angle \lambda_i$ tells roughly after how many generations the propagation may return to the original phasing but it does not mean returning to the original state since a transient state does not fail twice in one cascade.

The modal properties and their interpretations will be verified on the NPCC test system in the next section.

### E. Application of Modal Information

Modal properties of the stochastic interaction model can help prevent or reduce cascading failures. By reducing the probabilities of failure propagation at some vertices that highly participate in an unwanted mode, the mode can be weakened or even eliminated.

The vertices that highly participate in a mode can be identified by finding the eigenvector's entries with large magnitudes. For a persistent mode $(1, \mathbf{e}_k)$, eigenvector $\mathbf{e}_k$ indicates the only participation by $s_k$, which is verified by the self-loop at the vertex. For other modes, the highly participating vertices are identified from the corresponding eigenvector's entries that satisfy $|v_{ji}| \geq \varepsilon$. Here, $\varepsilon$ is a predefined threshold.

Considering that the vertices with trivial or persistent modes do not continue to propagate their failures, one mitigation strategy could transition those transient states corresponding to large, close-to-one diagonal elements of the interaction matrix into absorbing states. Such transitioning can be achieved by reducing the off-diagonal elements of their corresponding columns to zero, thereby enabling the associated cascades to end earlier. Actually, this mitigation strategy can be derived directly from the proposed stochastic interaction graph without eigen-analysis. Modeling multiple absorbing states in this paper allows for the distinction of cascades ending at different states. Since the eigen-analysis is our focus, the efforts for mitigation of cascading failures in this paper mainly focus on transient modes, especially those eigenvalues having large moduli since they can propagate to more vertices and their associated probabilities of failures may take several generations to decay to an ignorable value. For instance, by reducing the probabilities of key vertices with the largest positive less-than-one eigenvalue, failure propagation towards a wide area can be avoided. Similarly, transient modes with complex eigenvalues can also provide valuable information for mitigation of cascading failures. Transient modes with complex eigenvalues having nonzero angles form subgraphs, and the complexity of each subgraph relates to the eigenvalue's phase angle. A complex mode with a smaller angle often involves more vertices, indicating a more complex subgraph. If its modulus is close to one, it likely indicates a long-chain failure propagation, leading to a more severe cascade. Therefore, subgraphs formed by complex modes with moduli close to 1 and small angles are critical, indicating potential weak areas. This paper applies the nearly decomposable concept from [33] to mitigate cascading failures. Specifically, a mitigation strategy aims to reduce the connection between a critical subgraph and other states by decreasing the off-diagonal elements, i.e., the weights of links flowing into and out of this subgraph. Since fewer links flowing into the subgraph will result in fewer cascades propagating out through it, our mitigation mainly focuses on reducing the weights of links flowing into the subgraph. This ensures that states in a critical subgraph have weak interactions with other states. Consequently, the system state is less likely trapped into such weak areas, reducing the probability of large cascades.

## IV. NUMERICAL EXAMPLES

### A. Constructing a State-based Stochastic Interaction Graph

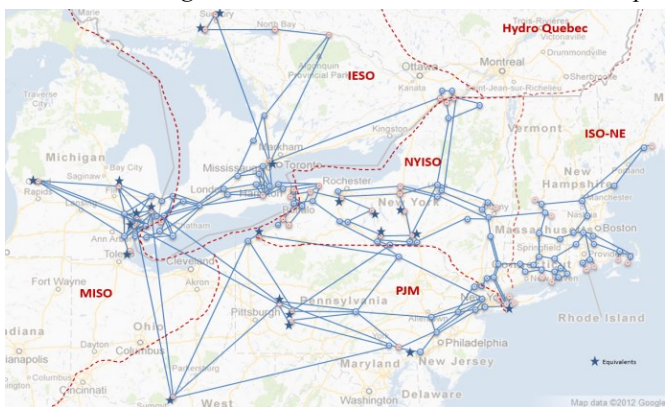

Fig. 4. NPCC 140-bus test system.

This section illustrates the construction of a stochastic interaction graph for failure propagation on a 140-bus NPCC test system as shown in Fig. 4 [43]. The system is a reduced model of the northeastern region power network, consisting of backbone transmission networks of ISO-NE, NYISO, PJM and some other northeastern regions. The ISO-NE region keeps the most details of the network [44]. To simulate cascading failures on the NPCC system, this study utilizes the improved OPA (ORNL-PSerc-Alaska) model in [8], which is based on a realistic steady-state power system model and considers both physical and control laws of power systems including power flows, line protection, load shedding and remedial actions. Starting from assumed initial failures, it can automatically generate cascades having sequential component outages grouped in generations. The OPA model and its variants have been widely used for cascading failure simulations to generate cascade datasets or validate mitigation strategies [7], [8], [14], [15], [18]. Its effectiveness in understanding the mechanism of failure propagations and finding statistical information on blackouts has been demonstrated well in the literature. For example, [45] shows that the outage data created by the OPA model on a 1553-bus system of the WECC (Western Electricity Coordinating Council) system align well statistically with historical blackouts in the actual WECC system.

The simulated failure data are used to construct the stochastic interaction graph by quantifying the interactions between component failures. Since the ISO-NE region is the most detailed region of the 140-bus system referred to the actual NPCC power network, all simulated cascades are assumed to originate from this region by randomly failing one of its 42 branches. In each cascade, failures can propagate from the New England region into the neighboring NYISO and PJM regions. Outside these three regions, the power network is highly reduced. Thus, terminate the simulation of each cascade once failures leave these three regions toward the external area. From the dataset, the quantification methods in Section II construct a state-based stochastic interaction graph having vertices and directed edges. The graph is denoted by $G_{NPCC}$ ($V_{NPCC}$, $E_{NPCC}$, $W_{NPCC}$) and characterized by a left stochastic matrix $\mathbf{W}_{NPCC}$.

Due to uncertainties of cascading failures, a sensitivity analysis of the stochastic interaction matrix is conducted. 15 cascading failure datasets having different numbers of cascades are generated on the NPCC system. We mainly focus on comparing transient modes of different datasets because persistent modes and trivial modes are characterized by constant unity and zero eigenvalues, respectively, and these modes do not continue to propagate their failures.

Fig. 5 shows how the eigenvalues of four transient modes having the largest moduli change with the increase of the dataset size. These modes are the top positive eigenvalue, top negative eigenvalue having the largest modulus, and top-2 complex eigenvalues respectively in (a), (b), (c), and (d) of Fig. 5. As the dataset size increases, these eigenvalues become more constant, especially when the size reaches 130k. Therefore, 130k is selected as the dataset size for case studies that follow.

The constructed state-based stochastic interaction graph has 672 vertices and 1527 directed edges, among which 325 edges are self-loops. The graph is visualized in Fig. 6. Although the theoretical dimensions of the model, such as the maximum number of vertices, may be huge, the actual dimension is reasonably limited as shown in Fig. 6. This is because, in practice, most subsets of failure components do not introduce associated vertices in the dataset. Thus, eigen-analysis is an efficient tool on the stochastic interaction model to understand patterns of failure propagation. The proposed eigen-analysis in Section III is conducted below to identify different types of modes on failure propagation with the associated stochastic interaction matrix. Interpretations of these modes are further verified on cascading failures of the system.

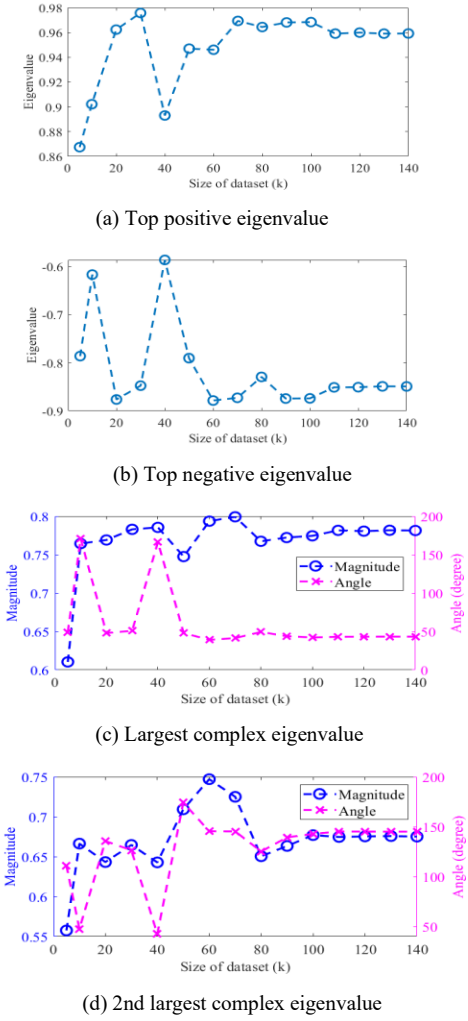

(a) Top positive eigenvalue

(b) Top negative eigenvalue

(c) Largest complex eigenvalue

(d) 2nd largest complex eigenvalue

Fig. 5. Eigenvalues of four transient modes that have the largest moduli based on the datasets of different sizes.

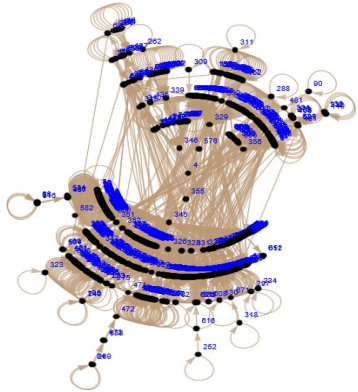

Fig. 6. Stochastic interaction graph.

### B. Identifying Different Modes of Failure Propagation

Eigen-analysis on the stochastic interaction model, derived from the entire cascade dataset, identifies 325 persistent modes, 285 trivial modes, and 62 transient modes. The modes are utilized to analyze the overall characteristics of cascading failure propagation based on the entire dataset, rather than focusing solely on a specific cascade. Among these transient modes, 10 modes have positive less-than-one eigenvalues, while 52 modes have complex eigenvalues with nonzero angles, including 8 eigenvalues with the angle of $\pi$, i.e., negative eigenvalues, and 22 pairs of conjugated eigenvalues. Key vertices in a mode can be identified via the orders of values with large amplitude in an eigenvector associated with that mode. The details are explained in subsection IV-C. In the following, the properties of all various modes are discussed and verified.

#### 1) Persistent Modes ($\lambda_i=1$)

The 325 persistent modes are associated with 325 absorbing states, respectively. These absorbing states make the Markov chain reducible, and the conclusions drawn in [33] for the reducible Markov chain can be applied here. Specifically, the states excluding absorbing states have zero entries in eigenvectors corresponding to all unity eigenvalues, indicating zero probability to achieve a steady state [33], or in other words, these states are transient states, not absorbing states. For instance, as shown in Fig. 7, mode $(1, \mathbf{e}_4)$ has a self-loop at vertex $s_4$, indicating the end of failure propagation at that state. There are also three edges coming from other three vertices to $s_4$: $e_{346,4}=(s_{346}, s_4)$, $e_{355,4}=(s_{355}, s_4)$, and $e_{576,4}=(s_{576}, s_4)$. When failures propagate to $s_4$ from any of these three other states, the propagation will end. The persistent mode associated with $s_4$ is verified: all 2217 out of 130k cascades that reach $s_4$ do end at it; particularly, 2216 cascades end in generation 1, and 1 cascade ends in generation 2.

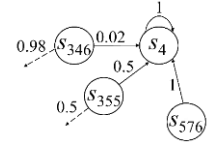

Fig. 7. Subgraph of a persistent mode.

#### 2) Trivial Modes ($\lambda_i=0$)

Totally, 285 trivial modes are identified. For each of 285 trivial modes, the highly participating vertices are either absorbing states or directly connected to the absorbing states, which verifies the fact that failures from the vertices highly participating in the mode last for no more than one generation. Fig. 8 illustrates two of the 285 trivial modes: $(0, \mathbf{v}_{560})$ and $(0, \mathbf{v}_{561})$. For the first mode, eigenvector $\mathbf{v}_{560}$ has $v_{171,560}=-0.5$, $v_{297,560}=-0.5$, $v_{586,560}=1$, and zeros for the other entries. A failure propagation reaching any of the three vertices will end at $s_{171}$ or $s_{297}$ by at most one more generation. Note that the two edges leaving $s_{586}$, $e_{586,171}=(s_{586}, s_{171})$ and $e_{586,297}=(s_{586}, s_{297})$, have a total probability of 1 because of $w_{171,586}=0.5$, $w_{297,586}=0.5$. Fig. 8 (b) shows another trivial mode $(0, \mathbf{v}_{561})$ having $v_{252,561}=-1$, $v_{616,561}=1$, and zeros for the others. A failure propagation reaching $s_{252}$ or $s_{616}$ will end at $s_{252}$.

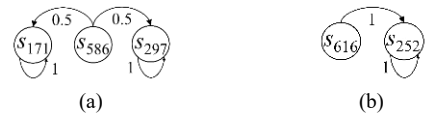

(a)          (b)

Fig. 8. Subgraphs of two trivial modes.

#### 3) Transient Modes ($0<|\lambda_i|<1$)

*(1) Positive Less-than-one Eigenvalues*

For a transient mode with a positive less-than-one eigenvalue, transient states with zero entries in its eigenvector indicate that failure propagation is unlikely to remain in these states after multiple generations. This suggests that these states will transition to an absorbing state and terminate early. A similar result can be observed in [33]. To verify this, select the transient mode (0.96, $v_{326}$) having the largest positive eigenvalue. According to the entries of $v_{326}$, transient states can be partitioned into two kinds: 87 entries having moduli less than $10^{-12}$ are considered rapidly decaying states and the other 260 vertices are considered slowly decaying states. For the other transient modes with positive less-than-one eigenvalues, such partitioning can also be done. However, since their eigenvalues are much smaller than 0.96, a failure propagation involving their non-zero entries may decay even faster. Among all cascades involving rapidly decaying states, only 0.01% of rapidly decaying states exist beyond two generations, verifying the fact that transient states with zero entries in its eigenvector help identify cascades that are less severe or more localized.

*(2) Complex Nonzero-angle Eigenvalues*

A complex eigenvalue having angle π is actually a negative eigenvalue. The transient mode with the smallest negative eigenvalue (i.e. the largest modulus) is (-0.85, $v_{327}$), denoted as the top negative mode. Consider the vertices with high participation in this mode, which are associated with $|v_{k,327}| \geq 0.5$ in $v_{327}$. Specifically, two 2-cycles involve four vertices $s_{352}$, $s_{357}$, $s_{358}$, and $s_{381}$ associated with $v_{352,327}=-0.77$, $v_{357,327}=-1$, $v_{358,327}=0.73$, and $v_{381,327}=0.9$ having high participations in the mode. One of the 2-cycles can be verified with two edges: $e_{352,381}=(s_{352}, s_{381})$ with $w_{381,352}=1$ and $e_{381,352}=(s_{381}, s_{352})$ with $w_{352,381}=0.3$, and the other 2-cycle can be verified with two edges: $e_{357,358}=(s_{357}, s_{358})$ with $w_{358,357}=0.6$ and $e_{358,357}=(s_{358}, s_{357})$ with $w_{357,358}=0.8$, as shown in Fig. 9.

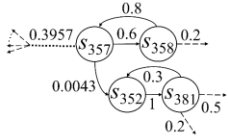

Fig. 9. Subgraph of a negative-eigenvalue mode.

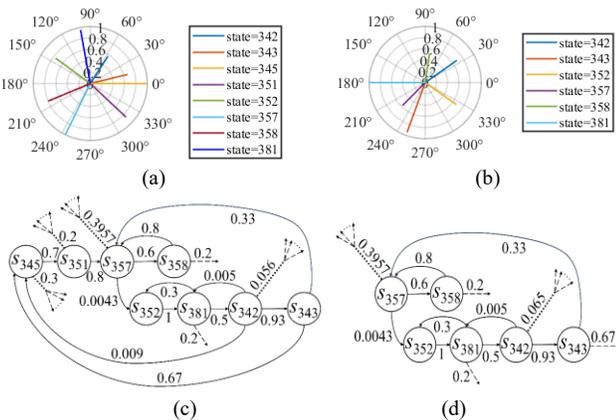

Fig. 10. Top-2 complex modes with mode shapes in (a) and (b) and subgraphs in (c) and (d).

Two pairs of complex conjugate eigenvalues having angle $\angle\lambda_i \neq \pi$ and the largest moduli are identified as transient modes. In the following discussion, we focus on the mode having $\angle\lambda_i \in (0, \pi)$ from each pair since its conjugate carries the same modal information. Name these two picked modes as the largest complex mode and the second largest complex mode. Highly participating vertices with $|v_{ji}| \geq 0.5$ are discussed.

Fig. 10 (a) shows the largest complex mode (0.78∠43.25°, $v_{328}$). It involves at least eight vertices to form a subgraph, which is determined by 360°/43.25°≈8. Its highly participating vertices are $s_{342}$, $s_{343}$, $s_{345}$, $s_{351}$, $s_{352}$, $s_{357}$, $s_{358}$, $s_{381}$. There are strong connections between $s_{342}$ and $s_{381}$, between $s_{345}$ and $s_{351}$, and between $s_{352}$ and $s_{381}$, based on the observation that angle differences $\angle v_{381,328}-\angle v_{342,328}$, $\angle v_{345,328}-\angle v_{351,328}$, and $\angle v_{352,328}-\angle v_{381,328}$ are all equal to 43.25°. These strong connections can be verified by corresponding edge weights shown in Fig. 10 (c).

Fig. 10 (b) shows the second largest complex mode (0.67∠145.05°, $v_{330}$). It involves at least 2 vertices to form a subgraph, which is determined by 360°/145.05°≈2. Its highly participating vertices $s_{342}$, $s_{343}$, $s_{352}$, $s_{357}$, $s_{358}$, $s_{381}$ form three more strongly connected subsets $\{s_{342}, s_{381}\}$, $\{s_{352}, s_{381}\}$, and $\{s_{357}, s_{358}\}$ based on the fact that angle differences $\angle v_{381,330}-\angle v_{342,330}$, $\angle v_{352,330}-\angle v_{381,330}$ and $\angle v_{357,330}-\angle v_{358,330}$ are all equal to 145.05°. The edges connecting the two vertices in each subset have relatively larger weights as shown in Fig. 10 (d): $w_{342,381}=0.5$ and $w_{381,342}=0.005$; $w_{352,381}=0.3$ and $w_{381,352}=1$; $w_{357,358}=0.8$ and $w_{358,357}=0.6$.

In observation, the number of vertices involved in a subgraph of each complex mode indicates the complexity of this complex mode. Specifically, the largest complex mode has greater complexity than the second largest complex mode from Fig. 10, based on the comparison of their mode shapes or subgraphs.

To summarize the interpretations on the different types of modes, persistent modes can reveal ending failures; trivial modes can identify sets of vertices that are unlikely to propagate further; transient modes indicate decaying probabilities with failure propagations. Particularly, different types of transient modes demonstrate different failure propagation characteristics. The transient mode with the largest positive eigenvalue can distinguish the slowly decaying states from the rapidly decaying states, which helps predict the severity of a cascade. Each transient mode with a negative eigenvalue involves two different vertices to form a 2-cycle, and each transient mode with a complex eigenvalue involves multiple vertices in a subgraph, telling to what extent failures under this mode may propagate and affect.

### C. Application of Modal Information

Consider 4 transient modes having the largest eigenvalue moduli: transient mode with the largest positive eigenvalue (0.96, $v_{326}$), transient mode with a negative eigenvalue (-0.85, $v_{327}$), the largest complex mode (0.78∠43.25°, $v_{328}$) and the second largest complex mode (0.67∠145.05°, $v_{330}$). Mitigation strategies will be designed and compared on these modes.

Highly participating transient states with the largest positive eigenvalue are identified from the largest $S$ entries (by modulus) of eigenvector $v_{326}$. $S'$ denotes the actual number of the associated transmission lines involved in the states with the top-$S$ entries, so there is $S' \geq S$.

Cascading failures can be mitigated by increasing transmission capacities of critical lines. In the following, we consider selective upgrading of transmission lines according to three strategies for comparison:

- **Strategy-R**: Upgrade transmission capacities of $S'$ randomly selected lines by 20%.

- **Strategy-Eigen**: Upgrade the transmission capacities by 20% for the *S'* lines associated with vertices highly participating in the largest positive eigenvalue.
- **Strategy-MF**: Upgrade the transmission capacities by 20% for the *S'* most frequently failed lines.

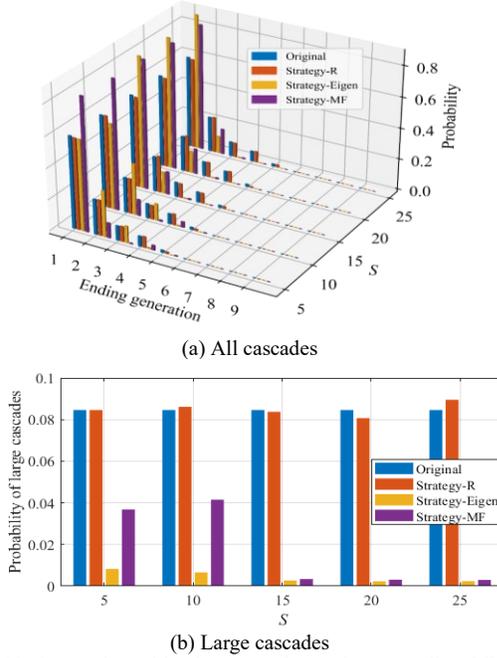

(a) All cascades

(b) Large cascades

Fig. 11. Comparing mitigation strategies against cascading failures.

After each strategy is implemented, the OPA model is used to re-generate a dataset of 130k cascades with the other parameters and settings unchanged. Fig. 11 (a) compares the probabilities of the ending generation in the original dataset and the datasets of three strategies with different $S$ values, each representing upgrading different numbers of transmission lines. $S=10$ corresponds to the entries greater than 0.1, and $S=25$ corresponds to the entries greater than 0.01. Fig. 11 (b) compares the probabilities of large cascades that propagate beyond generation 3. Table I shows the reduction in probabilities of large cascades after applying the three strategies. The dataset after random upgrading (Strategy-R) has a probability distribution for the ending generation similar to that of the original dataset without upgrading. Comparatively, intentional upgrading by the Strategy-Eigen or Strategy-MF can largely reduce the number of total generations to shorten failure propagations. Also, their performance depends on the value of $S$. Overall, a bigger $S$, i.e., upgrading more lines leads to better mitigation. When upgrading fewer lines ($S=5$ or $S=10$), the Strategy-MF performs better than the Strategy-Eigen in reducing small cascades that end in generation 2 or 3. However, as shown in Table I, Strategy-Eigen reduces the probability of large cascades by more than 33% compared to Strategy-MF. This demonstrates that Strategy-Eigen has an obvious advantage over Strategy-MF in reducing the probabilities of large cascades. After Strategy-Eigen, almost all cascades end before or at generation 3. Compared with the large cascades, the small cascades have much less impact on the system, and they can be addressed in the system planning stage by contingency analyses and protection designs. When upgrading a large number of lines ($S\geq15$), the Strategy-Eigen is better than the Strategy-MF since the former results in a lower probability for any cascade beyond generation 1. Therefore, a mitigation strategy designed based on the largest positive eigenvalue is more effective in shortening failure propagations than a strategy based on the most frequent failures.

TABLE I
REDUCTION IN PROBABILITIES OF LARGE CASCADES AFTER APPLYING DIFFERENT STRATEGIES

| Strategy \ S | Strategy-R | Strategy-Eigen | Strategy-MF |
|---|---|---|---|
| 5 | 0% | 90.2% | 56.5% |
| 10 | -1.8% | 92.2% | 51.0% |
| 15 | 1.1% | 96.8% | 96.0% |
| 20 | 4.6% | 97.3% | 96.3% |
| 25 | -5.6% | 97.2% | 96.4% |

As discussed in subsection III-E, mitigation strategies can also be designed based on the complex eigenvalues having nonzero angles. As illustrated in subsection IV-B, the largest complex mode is characterized by the eigenvalue with a greater modulus and a smaller angle compared with that of the second largest complex mode. Also, the subgraphs identified by the top negative mode and the second largest complex mode all highly participate in the largest complex mode. Therefore, the mitigation strategy focuses on nearly decomposing the subgraph of the largest complex mode from the other states. The sum of weights of the links that inject from the other states outside the subgraph into its vertices, denoted by $S_w$, indicates the interactions among the subgraph and the other states. The mitigation strategy should focus on weakening these interactions to make this subgraph nearly decomposable from other states. A vertex with a bigger $S_w$ indicates its stronger interaction with the outside states, and this vertex is more critical for mitigating cascading failures. In the subgraph, the vertex $s_{345}$ has the largest $S_w$, accounting for 60% of the total $S_w$ of all vertices in the subgraph. Therefore, vertex $s_{345}$ is the one having the most significant interactions with the outside states. Two mitigations similar to Strategy-Eigen are implemented here to increase the transmission capability of lines in $s_{345}$ and all states in the subgraph by 20%, respectively. Then, two new datasets of 130k cascades are re-generated. After counting the large cascades caused by all vertices in the subgraph, their total number is reduced by 89.9% and 95.8%, respectively, demonstrating the value of knowing complex modes in designing mitigation strategies.

## V. CONCLUSIONS

In this study, a stochastic interaction graph model is proposed to better quantify the interactions between component failures on a power system. The characteristics of failure propagation beyond the physical topology of the power system are explored through innovative eigen-analysis on the proposed stochastic interaction graph as a directed graph. Different types of failure propagation modes including persistent modes, trivial modes, and transient modes, characterized by eigenvalues, whose absolute values are respectively unity, zero, and in between, are defined and analyzed. The proposed stochastic interaction graph and matrix, eigen-analysis, and interpretations of modes are verified using datasets of simulated cascading failures on an NPCC test system. Accordingly, mitigation strategies have been designed and compared to show the value of modal information in failure propagation.

## VI. Acknowledgement

The U.S. Government retains the publisher, by accepting the article for publication, and acknowledges that the United States Government retains a non-exclusive, paid-up, irrevocable, world-wide license to publish or reproduce the published form of this manuscript, or allow others to do so, for the U.S. Government purposes. The Department of Energy (DOE) will provide public access to these results of federally sponsored research in accordance with the DOE public access plan.

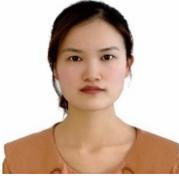
**Zhenping Guo** (Student Member, IEEE) received her B.S. and M.S. degrees in electrical engineering from Wuhan University of Technology, Wuhan, China, in 2013, and Wuhan University, Wuhan, China, in 2016, respectively. She is currently pursuing a Ph.D. degree with the Department of Electrical Engineering and Computer Science, University of Tennessee, Knoxville, TN, USA. From 2016 to 2021, she was an Engineer with the State Grid Hubei Power Supply Company. Her research interests include cascading failures, power system simulations, and transient stability analysis.

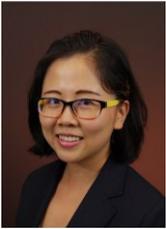
**Xiaowen Su** (Member, IEEE) has been an Assistant Profession of Research in the Department of Electrical and Computer Engineering (ECE) at the University of Texas at San Antonio (UTSA) since 2022. Her research interests include power system cascading failure analysis, frequency estimation, voltage stability index, and power system model reduction. Prior to joining UTSA, she worked as a Research Associate in the Department of Electrical Engineering and Computer Science (EECS) at the University of Tennessee at Knoxville (UTK). She earned her Ph.D. in 2019 from the Department of Mechanical, Aerospace and Biomedical Engineering (MABE) at UTK, specializing in Vibration and Control.

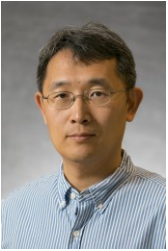
**Kai Sun** (Fellow, IEEE) is a professor with the Department of Electric Engineering and Computer Science, University of Tennessee, Knoxville, USA. He received the B.S. degree in automation in 1999 and the Ph.D. degree in control science and engineering in 2004 both from Tsinghua University, Beijing, China. Before he joined the University of Tennessee, he was a Project Manager in grid operations, planning and renewable integration with the Electric Power Research Institute (EPRI), Palo Alto, CA.

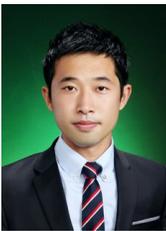
**Byungkwon Park** (Member, IEEE) received the B.S. degree in electrical engineering from Chonbuk National University, South Korea, in 2011, and the M.S and Ph.D. degrees in electrical engineering from the University of Wisconsin-Madison, Madison, WI, USA, in 2014 and 2018, respectively. He is currently an assistant professor with the School of Electrical Engineering at the Soongsil University (SSU), Seoul, South Korea. Before joining SSU, he was a Research Staff with the Oak Ridge National Laboratory, Computational Sciences and Engineering Division. His research interests include modeling, simulation, control, and optimization of electrical energy systems.

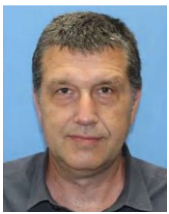
**Srdjan Simunovic** received the M.Sc. and Ph.D. degrees in civil engineering from Carnegie Mellon University, Pittsburgh, PA, USA. He is a Distinguished Research Staff with the Computational Sciences and Engineering Division, Oak Ridge National Laboratory, Oak Ridge, TN, USA. His research expertise includes applied mathematics for multi-scale and multi-physics problems, parallel algorithms and computing, energy storage, and computational modeling of materials and processes.